\documentclass[11pt]{article}
\usepackage[pdftex]{graphicx}

\usepackage{cite}
\usepackage{color}
\usepackage{amsfonts}
\usepackage{amsmath}
\usepackage{physics}
\usepackage{latexsym}
\usepackage{fancybox}
\usepackage{subfigure}
\usepackage{amssymb}
\usepackage{cite}
\usepackage{url}
\usepackage{here}
\usepackage{mathtools}
\numberwithin{equation}{section}
\allowdisplaybreaks

\def\spa#1{\phantom{\fbox{\rule[-#1cm]{0cm}{0cm}}}}

\def\[#1\]{\begin{align}#1\end{align}}
\begin{document}

\hfuzz=100pt
\title{{\Large \bf{
Pinched geometries in $\mathbf{2}$D Lorentzian quantum Regge calculus
}}}
%\date{}
\author{
Yoshiyasu Ito$^{a}$\footnote{ito@eken.phys.nagoya-u.ac.jp},
Daisuke Kadoh$^{b}$\footnote{kadoh@mi.meijigakuin.ac.jp}, 
Yuki Sato$^{cd}$\footnote{yukisato@u-fukui.ac.jp}
  \spa{0.6} \\
\\
$^a${\small{\it Market Finance Department, The Ogaki Kyoritsu Bank, Ltd.}}
\\ {\small{\it 3-98 Kuruwamachi, Ogaki-shi, Gifu 502-0887, Japan}}\\
\\
$^b${\small{\it Institute for Mathematical Informatics, Meiji Gakuin University}}
\\ {\small{\it Yokohama-shi, Kanagawa 244-8539, Japan}}\\
\\
$^c${\small{\it Department of Mechanical and System Engineering, University of Fukui}}
\\ {\small{\it 3-9-1 Bunkyo, Fukui-shi, Fukui 910-8507, Japan}}\\    
\\
$^d${\small{\it Department of Physics, Nagoya University}}
\\ {\small{\it Chikusaku, Nagoya-shi, Aichi, 464-8602, Japan}}\\
\spa{0.3} 
}
\date{}

\maketitle
\centerline{}

\begin{abstract} 

We investigate pinched geometries in a two-dimensional Lorentzian model of quantum Regge calculus (QRC) using the tensor renormalization group (TRG) method. 
A pinched geometry refers to a configuration with an infinitely long temporal extent,
even when the total spacetime area is fixed.   
We examine several choices of  integration measures and triangulations to study
whether such geometries can dominate in the limit of infinitely many triangles. 
Our results indicate that pinched geometries are strongly suppressed, and 
 this suppression is observed across different integral measures and triangulations. 
These results suggest the possible emergence of smooth geometries as well as a sort of universality for infinitely many triangles.

\end{abstract}

\renewcommand{\thefootnote}{\arabic{footnote}}
\setcounter{footnote}{0}

\newpage

\section{Introduction}
\label{sec:Introduction}

Regge's idea of discretizing spacetime using simplices \cite{Regge:1961px} has been extended 
to Quantum Regge Calculus (QRC), which is a lattice model of quantum gravity (see e.g. Refs.~\cite{Williams:1991cd, Williams:1996jb, Hamber2007, Barrett:2018ybl}). 
This formulation assumes  a fixed triangulation and treats the edge lengths of each simplex as dynamical variables.
The Euclidean model has been extensively studied so far (see e.g., Refs.~\cite{Hamber:1984kx, Hamber:1985qj, Hamber:1985gw, Hamber:1985gx, Jevicki:1985ta, Nishimura:1994qg}), but it is known to suffer from singular configurations called spikes, 
for which various discussions have been presented in the literature \cite{Bock:1994mq, Holm:1995kw, Ambjorn:1997ub,Rolf:1998ja}.
In contrast, the $2$D Lorentzian QRC model forbids such spike configurations \cite{Tate:2011ct, Jia:2021deb}.
\footnote{
Singular configurations in $3$D and $4$D Lorentzian QRC have been extensively studied in Refs.~\cite{Borissova:2024pfq, Borissova:2024txs}. 
}
However, numerical approaches to the Lorentzian model remain challenging because of the sign problem.

The TRG method is a promising and rapidly developing way for numerical studies of models with the sign problem. 
This method was first introduced by Levin and Nave in the $2$D Ising model \cite{Levin:2006jai},  
and some improvements and higher-dimensional extensions were also studied in 
Refs.~\cite{Evenbly:2015ucs, Yang:2017lvo, Xie2012, Adachi:2019paf, Kadoh:2019kqk}.
Unlike the Monte Carlo method, the TRG does not require any statistical procedures, and the sign problem does not arise in the first place. This method also enables relatively low computational cost for large lattice volumes.
Exploiting these benefits, the TRG has already been applied to lattice field theory computations, e.g., 
Refs.~\cite{Liu:2013nsa, Shimizu:2014uva, Zou:2014rha, Shimizu:2014fsa, Takeda:2014vwa, Yang:2015rra, Sakai:2017jwp, Kadoh:2018hqq,  Kadoh:2018tis, Bazavov:2019qih, Meurice:2019ddf, Asaduzzaman:2019mtx, Butt:2019uul, Kadoh:2019ube, Akiyama:2020ntf, Akiyama:2020sfo, Bloch:2021mjw, Akiyama:2022eip,Yosprakob:2023tyr, Asaduzzaman:2023pyz,Yosprakob:2023jgl} (See Ref.~\cite{Meurice:2020pxc, Kadoh:2022loj} for recent developments)
and is expected to be useful for studying the quantum universe composed of a large number of simplices.\footnote{
See Refs.~\cite{Dittrich:2014mxa, Delcamp:2016dqo, Cunningham:2020uco} for the pioneering works on application of TRG to spin foam models that are closely related with QRC.}

In our previous article \cite{Ito:2022ycc}, we applied TRG to a $2$D Lorentzian QRC model proposed by Tate and Visser \cite{Tate:2011ct}.   
We showed some evidence that such configurations would be suppressed as the number of lattice sites goes to infinity.
The suppression of pinched geometry implies that smooth geometries may emerge from the $2$D Lorentzian QRC. 

To make our conclusion about the pinched geometry more convincing, further studies are required. In Ref.~\cite{Ito:2022ycc}, 
we numerically demonstrated that the mean squared length of time-like edges is finite
for a fixed regular triangulation with a particular choice of integral measure. 
As a next step, we plan to study the higher moments of the time-like edge length for a few integral measures and triangulations. 
Such studies would also allow us to check whether the suppression of pinched geometries can be universal or not. 

The rest of this article is organized as follows.       
In Sec.~\ref{sec:QRC} we explain $2$D Lorentzian QRC in detail, and show how to express the model in terms of tensor networks. Sec.~\ref{sec:Results} presents the numerical results obtained in this work. The detailed numerical studies strongly suggest the suppression of pinched geometries for a variety of integral measures and triangulations. 
Finally, Sec.~\ref{sec:Discussions} is devoted to discussions.

\section{$\mathbf{2}$D Lorentzian QRC and tensor networks}
\label{sec:QRC}

Lorentzian quantum Regge calculus (QRC) is a simplicial model of quantum geometries.  
In this section, we review $2$D Lorentzian QRC \cite{Tate:2011ct} and 
its tensor network representation \cite{Ito:2022ycc} with a different triangulation. 

\subsection{$\mathbf{2}$D Lorentzian QRC}
\label{sec:2dLQRC}
\begin{figure}[h]
\centering
\includegraphics[width=4.0in]{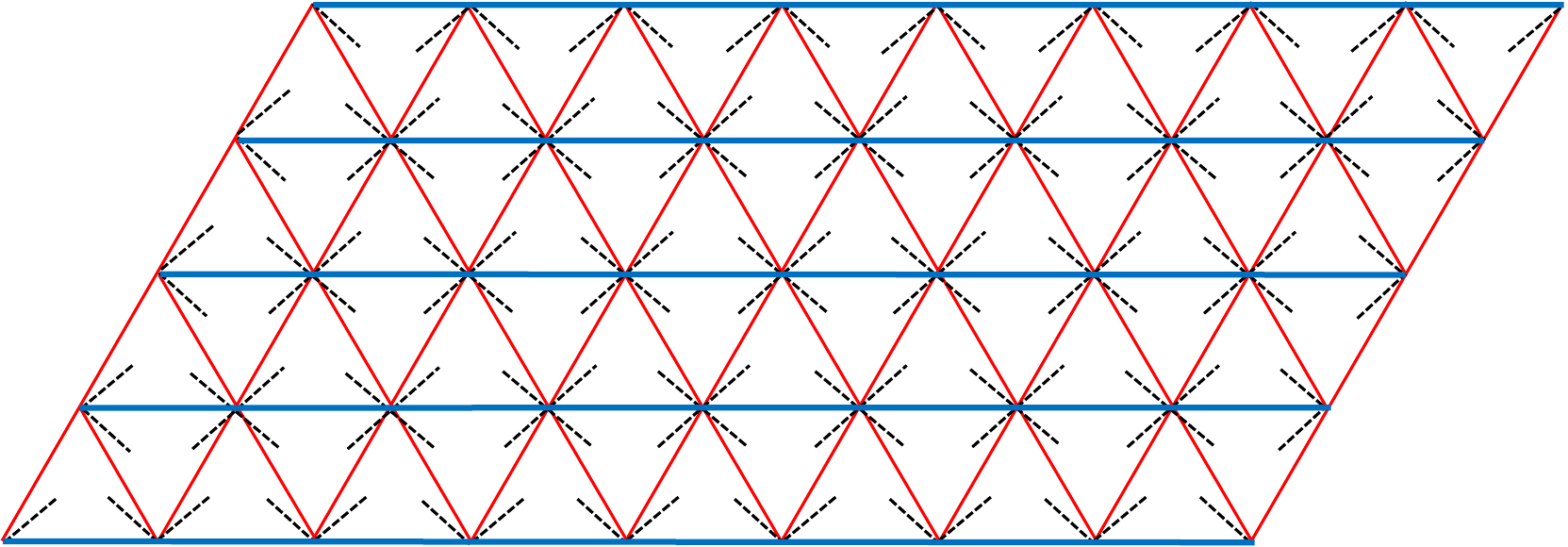}
\caption{A regular triangulation with a time foliation and a single light-cone at each vertex: Thick blue lines, thin red lines and dashed black lines indicate space-like edges, time-like edges and light rays, respectively.}
\label{fig:triangulation}
\end{figure}

$2$D QRC is a lattice theory that is defined on a fixed triangulation such that the edge lengths of each triangle are dynamical variables.  
A Lorentzian model introduced by Tate and Visser \cite{Tate:2011ct} does not suffer from the problem of spike configurations that appear in the Euclidean model. 
In this section, we specifically discuss a special class of regular triangulations with a time foliation equipped with exactly four light rays at each vertex (see Fig.~\ref{fig:triangulation}), as used in Ref.~\cite{Tate:2011ct}. 
This paper focuses on the torus topology, but its generalization to other topologies would be straightforward.

\begin{figure}[h]
\centering
\includegraphics[width=3.0in]{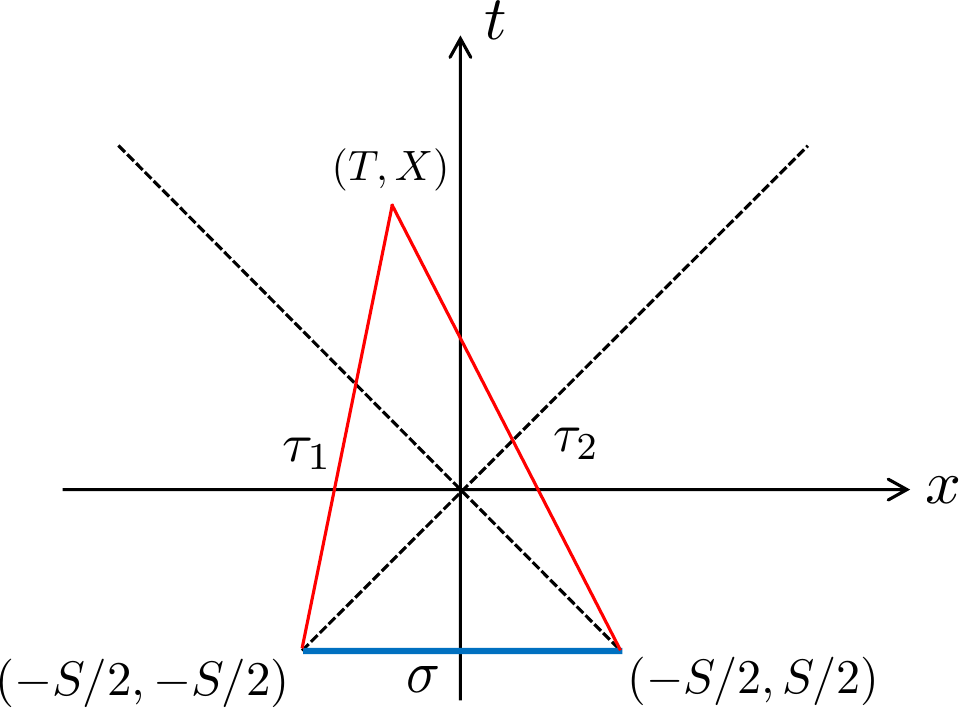}
\caption{A triangle defined in the $2$D Minkowski space.}
\label{fig:lorentziantriangle}
\end{figure}

In the Lorentzian model, every point in a triangle (including edges and vertices)
can be specified by the local Minkowskian coordinates. 
Therefore, the invariant distance between two vertices 
distinguishes the type of edges as space-like or time-like.
In the Tate-Visser type Lorentzian QRC, a specific triangle with one space-like edge and two time-like edges 
is used as a building block (see Fig~\ref{fig:lorentziantriangle}). 
Here, $\tau_1$ and $\tau_2$ are the time-like edge lengths (proper times) and $\sigma$ is the space-like edge length (proper distance). 
In Fig.~\ref{fig:lorentziantriangle}, the coordinates of the three vertices are 
$(-S/2,-S/2)$, $(-S/2,S/2)$ and $(T,X)$. 
The area of the triangle is therefore given by 
$A=\frac{S}{2}\left(\frac{S}{2} + T \right)$,
which can also be expressed using the edge lengths as
\[
A (x,y,z) 
= \frac{1}{4} \sqrt{ x^2 + y^2 +z^2 +2z (x+y) -2 xy}\ ,
\label{eq:area}
\]
where $x=\tau_1^2$, $y=\tau_2^2$ and $z=\sigma^2$, and the inner products between two vectors are defined with respect to the Minkowski metric, $\eta = \text{diag}(-1,1)$. 
The Euclidean case requires the triangle inequalities 
to form a regular triangle. However, such extra constraints are not necessary in the $2$D Lorentzian case  
because $A(x,y,z)$ is positive for given three edges satisfying $x, y \ge 0$ and $z>0$.

Let $N$ be the number of triangles and $N_e$ be the number of edges in the given triangulation. 
Note that $N$ is an even number, $N_e$ is given by $N_e=3N/2$, the number of space-like edges is $N/2$ and that of time-like edges is $N$.  
We then label each triangle and let $A_s$ be the area of the $s$-th triangle where $s=1, 2.\cdots, N$. 
We also label the edges whose lengths are given by $\ell_e\, (e=1, 2,\cdots,N_e)$.

Since the scalar curvature term in the $2$D Einstein-Hilbert action becomes a constant known as the Euler characteristic, it does not play any role if fixing the topology.   
We then introduce the cosmological constant term as the gravitational action whose discretization is given by the summation of triangle areas in the triangulation:
\[
S = \lambda \sum_{s=1}^N A_s[\{\ell^2\}]\ , 
\label{eq:Reggeaction}
\]
where $\lambda$ is the cosmological constant.  
The lattice action (\ref{eq:Reggeaction}) is called the Regge action, in which the edge lengths are treated as dynamical variables. 

The path integral is given by the multiple integral over the domain of edge lengths: 
\[
Z
= \int \, d\mu[\{\ell^2\}]  \ 
e^{i S[\{\ell^2\}]}\ .
\label{eq:Z}
\]
The integral measure  $d\mu[\{\ell^2\}]$ is not uniquely determined and in general can be non-local\footnote{
A discretization-invariant local path integral measure for $3$D Lorentzian QRC was proposed in Ref.~\cite{Borissova:2023izx}.
}.  
In this article, we study two types of local measure that are analogous to the ones commonly used in the Euclidean case (see e.g. Ref~\cite{Hamber:1991sy} and references therein): 
\[
d\mu[\{\ell^2\}] 
\equiv 
 \prod_s \left( A_s[\{\ell^2\}] \right)^\beta \cdot \prod_{e}  \ell^{2\alpha}_e d\ell^2_e, 
\label{eq:measure_beta}
\]
where $\alpha$ and $\beta$ are parameters and the multiple integration is performed over the domain $\ell_e^2 \in (0,\infty)$. 
We refer to the case with $\alpha \neq 0$ and $\beta = 0$ as the $\alpha$-measure, and the case with $\alpha = 0$ and $\beta \neq 0$ as the $\beta$-measure.

As discussed in Ref.~\cite{Ito:2022ycc}, the factor $iS$ in Eq.~\eqref{eq:Z} can be transformed into $-S$ by an analytic continuation that leaves both the definition of the triangle area (\ref{eq:area}) and the integration measure (\ref{eq:measure_beta}) unchanged. 
This is achieved by rewriting the variables $\ell_e^2$ in terms of the $N_e$-dimensional spherical coordinates $(r, \Omega_{N_e-1})$ and then rotating the integration contour of the radial variable $r$ to the imaginary axis. 
For more details, please refer to App.~B in our previous paper \cite{Ito:2022ycc}. 
We may consider the following partition function instead of  Eq.~\eqref{eq:Z}: 
\[
Z^\prime
= \int \, d\mu[\{\ell^2\}]  \ 
e^{- S[\{\ell^2\}]}\ ,
\label{eq:Z2}
\]
and the fixed-area partition function:\[
Z^\prime_A = 
\int d\mu[\{\ell^2\}] \ 
 \delta \left(\sum_s A_s [\{ \ell^2 \}] - A\right) 
 \ . 
\label{eq:Za} 
\]
If $\frac{3}{2}(\alpha + 1) + \beta > 0$, one can show that $Z'_A$ is proportional to $Z'$: 
\[
Z^{\prime}_A
=
 \frac{ (\lambda A)^{\theta} }{ A \Gamma (\theta) }\ Z^\prime \ , \ \ \ 
 \text{with}\ \ \ \theta := N \left( \frac{3}{2} (\alpha + 1) + \beta \right)\ . 
 \label{eq:relationzza}
\]
Hereafter we choose the integral measure in such a way as to satisfy the inequality $\frac{3}{2}(\alpha + 1) + \beta > 0$.

We introduce the expectation value of an operator ${\cal O}$:
\[
\left\langle \mathcal{O} \right\rangle 
= \frac{1}{Z^\prime} 
\int d\mu[\{\ell^2\}]\ \mathcal{O} (\{\ell^2\}) \ 
e^{-S[\{\ell^2\}]}\ ,
\label{eq:ev}
\]
and the one with the area fixed: \[
\left\langle \mathcal{O} \right\rangle_A 
= 
\frac{1}{Z^\prime_A} 
\int d\mu[\{\ell^2\}] \ \mathcal{O} (\{\ell^2\}) \ 
 \delta \left(\sum_s A_s [\{ \ell^2 \}] - A\right) \ .
\label{eq:ev_A}
\] 
In fact, the expectation value of the area of a single triangle can be calculated exactly: 
\[
\left\langle A_s \right\rangle 
= 
- \frac{1}{N} \frac{d}{d\lambda} \log Z^{\prime} 
= \left( \frac{3}{2} (\alpha + 1) + \beta \right) \frac{1}{\lambda}\ . 
\label{eq:ev_area} 
\]
This quantity can serve as a useful benchmark for numerical calculations. 

Let $\mathcal{O}_m$ be an operator satisfying $\mathcal{O}_m (\{ \gamma \ell^2 \}) = \gamma^m \mathcal{O}_m ( \{ \ell^2 \} )$ for $\gamma>0$. 
Doing a little math yields  
\[
\left\langle \mathcal{O}_m \right\rangle_A 
= \frac{(\lambda A)^m \Gamma \left(  \theta  \right)} {\Gamma \left( \theta + m \right)}
\left\langle \mathcal{O}_m \right\rangle 
\ .
\label{eq:relation}
\]

In fact, the area function \eqref{eq:area} has a flat direction. For instance, $A$ remains unchanged
when $x=y=a$ and $z=\sqrt{4a^2+A^2}-2a$ for any value of $a$. 
In Ref.~\cite{Ito:2022ycc}, we introduced an infrared cutoff $\mu$ as a regulator for the flat direction. 
However, we have confirmed that at large $N$ numerical calculations can be done to good accuracy without introducing $\mu$.
Therefore, we do not introduce $\mu$ in this article and perform numerical calculations at large $N$.

\subsection{Spike and pinched geometry}

\begin{figure}[h]
\centering
\includegraphics[width=4.0in]{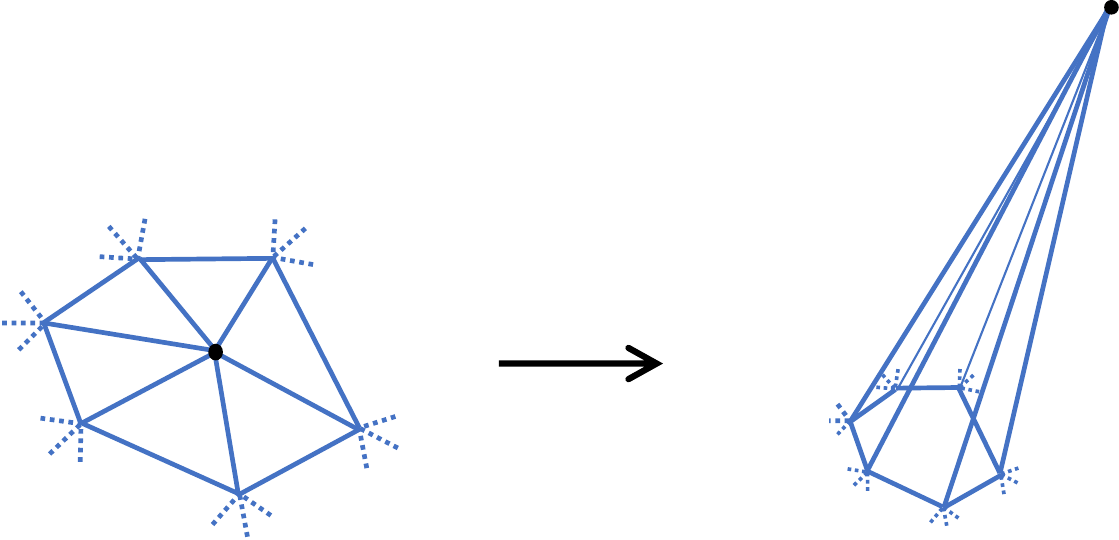}
\caption{An illustration of the spike.}
\label{fig:spike}
\end{figure}

$2$D Euclidean QRC has a singular configuration known as the spike, 
in which all the edge lengths emanating from a vertex are infinite while keeping the area of triangulation finite. 
Fig.~\ref{fig:spike} illustrates a spike configuration.    
It is actually shown that the area-fixed expectation value $\left\langle \ell^n \right\rangle_A$ diverges for sufficiently large positive value of $n$ \cite{Ambjorn:1997ub}. 
The smallest $n$ with  $\left\langle \ell^n \right\rangle_A=\infty$ depends on the measure and the triangulation method.  

In $2$D Euclidean QRC without the $R^2$ term, 
spike configurations emerge, causing the effective lattice spacing to diverge and preventing the theory from describing a smooth spacetime.

\begin{figure}[h]
\centering
\includegraphics[width=3.5in]{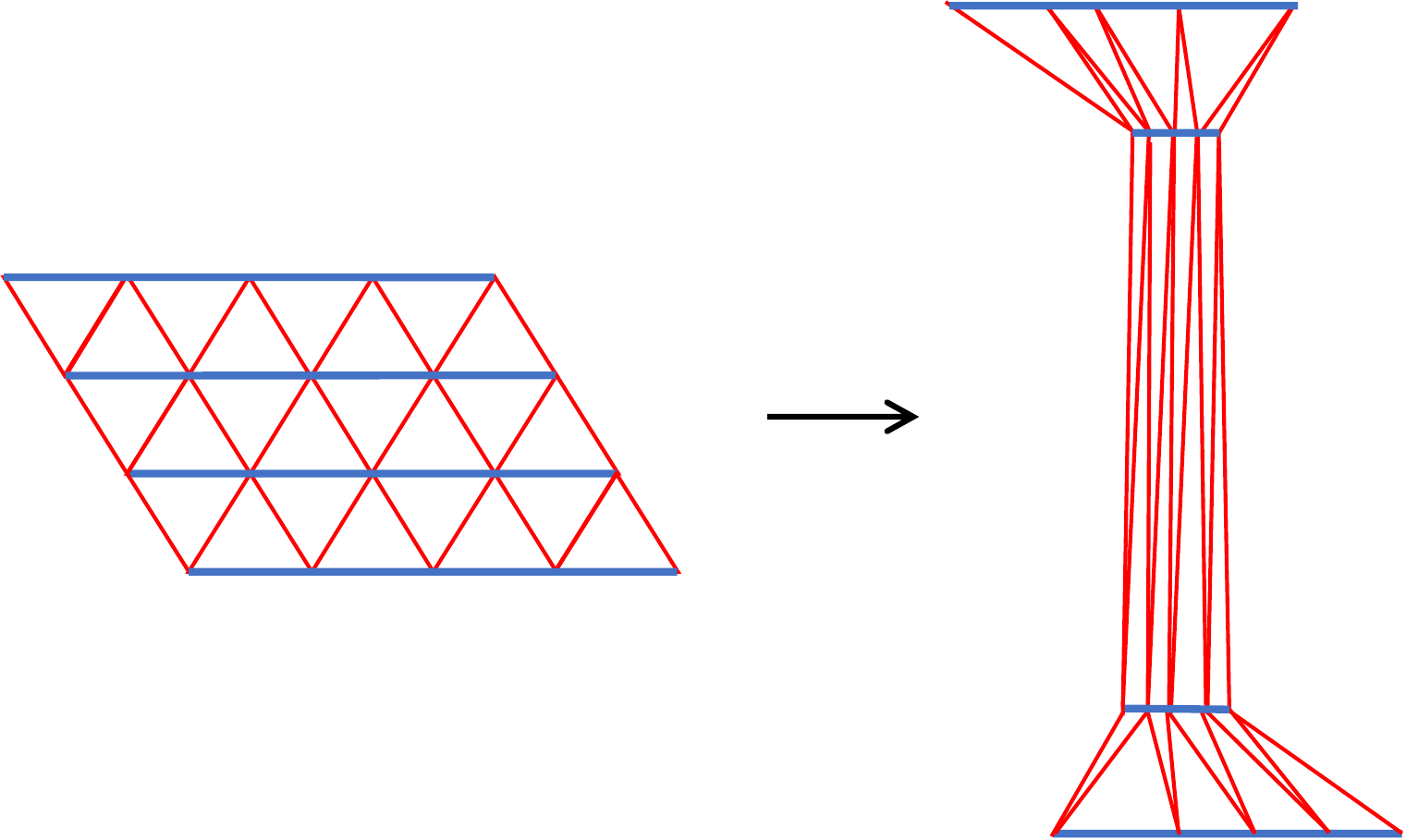}
\caption{An illustration of the pinched geometry.}
\label{fig:pinch}
\end{figure}

As shown in Refs.~\cite{Tate:2011ct,Jia:2021deb}, the Lorentzian model does not have this problem 
because the spike configuration is forbidden if a single light cone is placed at each vertex. If there were a spike, no light ray would emanate from the tip of a spike \cite{Jia:2021deb}.  
For any positive finite value of $n$, we have at large $N$,  
\[
\left\langle \sigma^n \right\rangle_A
< \infty\ . 
\label{eq:nospike}
\]
However,  the Lorentzian model still has a configuration with the divergent time-like edges as seen in Fig.~\ref{fig:pinch}. 
Such a configuration is called the pinched geometry. 
The question is whether the pinched geometries are dominant or not at large $N$. 
This might be checked through the investigation of the finiteness of     
$
\left\langle \tau^n \right\rangle_A
$ at large $N$.  

In our previous paper \cite{Ito:2022ycc}, 
we numerically confirmed the suppression of pinched geometries through the finiteness of $\left<\tau^2\right>_A$ for a regular triangulation shown in Fig.~\ref{fig:triangulation}. 
For different triangulations and integral measures, we verify the suppression by examining the higher moments $\left<\tau^n \right>_A$ up to $n=10$ in Sec.~\ref{sec:Results}.

\subsection{Tensor networks}
\begin{figure}[h]
\centering
\includegraphics[width=3.0in]{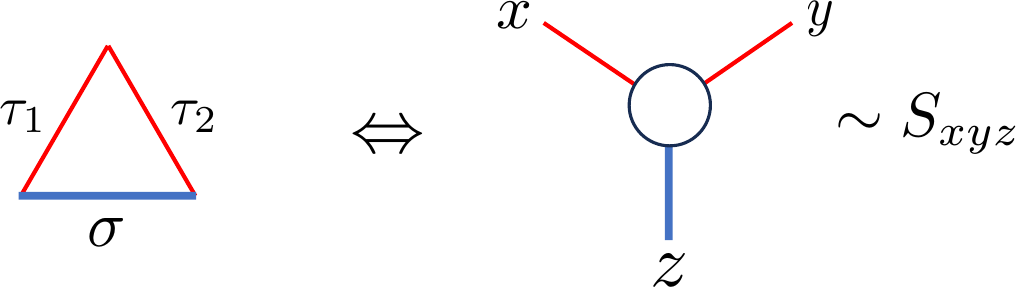}
\caption{A dual graph of a Lorentzian triangle: Each triangle with the edge lengths, $\sigma$, $\tau_1$ and $\tau_2$, in the triangulation 
corresponds to a trivalent vertex whose edges respectively carry continuous indices, $z=\sigma^2$, $x=\tau^2_1$ and $y= \tau^2_2$, in the dual picture, 
and the rank-$3$ tensor $S_{xyz}$ is assigned to each trivalent vertex.}
\label{fig:dualgraph}
\end{figure}

\begin{figure}[h]
\centering
\includegraphics[width=3.0in]{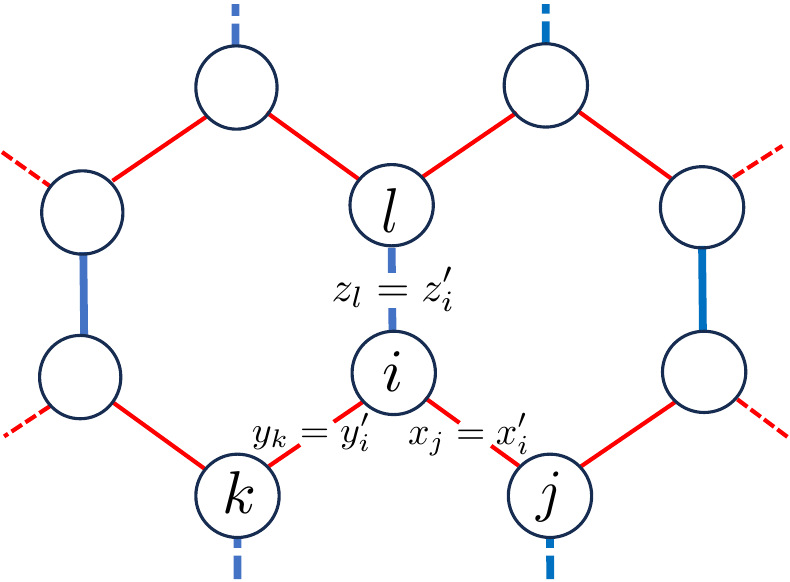}
\caption{Identification of the tensor indices.}
\label{fig:network}
\end{figure}

The partition function (\ref{eq:Z2}) can be expressed as a tensor network consisting of rank-$3$ tensors,
whose indices correspond to the squared edge lengths.
We define the rank-$3$ tensor carrying continuous indices as (see Fig.~\ref{fig:dualgraph}):
\[
S_{x y z} =  (xyz)^{\alpha/2} \left[ A (x,y,z) \right]^{\beta} \, e^{-\lambda A(x,y,z)}\ . 
\label{eq:tensor}
\]
Note that the tensor depends on the choice of integral measure. 
The partition function is then given by
\[
& Z^\prime
=
\int^{\infty}_{0} 
\prod_{i \in u}
dx_i dy_i dz_i 
\prod_{i \in u, j \in d}
S_{x_i, y_i, z_i}
S_{x^\prime_j y^\prime_j z^\prime_j} \ , 
\label{eq:ctn}
\]
where $u$ and $d$ respectively denote the sets of trivalent vertices corresponding to the upward and downward triangles, 
and the primed and unprimed indices are identified when the associated links are connected, as shown in Fig.~\ref{fig:network} (see Ref.~\cite{Ito:2022ycc} for details). 
With this identification rule, each index is shared by two tensors.
Thus, the integrals over indices correspond to tensor contractions,
and Eq.~\eqref{eq:ctn} can indeed be expressed in terms of a tensor network with continuous indices. 

The indices of the resulting tensor network have infinite dimension, and a finite-dimensional approximation is required for numerical computations. 
In the next section, we rewrite the tensor network with finite bond dimension by discretizing the integrals with quadrature methods and truncating the resulting sums.

\section{Numerical Results}
\label{sec:Results}

We investigate the finiteness of $\left<\tau^n\right>_A$ for various integration measures and values of $n$. 
In our previous work\cite{Ito:2022ycc}, we examined the finiteness of $\left<\tau^2\right>_A$ for the regular triangulation shown in Fig.~\ref{fig:triangulation}. 
However,  higher moments may potentially diverge.
To address this issue carefully, we extend our analysis up to $n=10$. 
As each expectation value varies smoothly and monotonically with respect to $n$, 
we mainly present the results for $n=2$ and $n=10$.
Furthermore, to explore the universality of our findings across different triangulations,  
we perform a similar analysis on the $8$-$4$ irregular triangulation (as shown in Fig.~\ref{fig:4-8_triangulation}) 
for various measures parametrized by $\alpha$ and $\beta$.  
Hereafter we set $\lambda = 1$ without loss of generality. 

\subsection{Discretization of tensor indices}

For numerical computations, we employ the higher-order tensor renormalization group (HOTRG) method \cite{Xie2012}. 
Hereafter we use $D$ for the cut for the bond dimension of the tensors in HOTRG. 
Details of how the HOTRG method is applied to the $2$D Lorentzian QRC can be found in our previous article \cite{Ito:2022ycc}. Our numerical procedure is essentially the same as in our previous work, except that the quadrature rule used to make the tensors finite-dimensional has been changed from the naive Gauss-Laguerre quadrature to the generalized Gauss-Laguerre quadrature. This is because generalized one is more effective for the $\alpha$-measure with $\alpha\neq 0$, as seen later.

The bond dimension of tensors must be finite in the TRG computations. 
To this end, we employ the generalized Gauss-Laguerre quadrature method: 
\begin{align}
  \int_0^\infty \dd x \, x^a e^{-x} f(x) \simeq
  \sum_{x\in S_K} w_K(a,x) f(x)\ , \ \ \  (a>-1)
\end{align}
where $S_K = \{ x_i | L_K^{(a)}(x_i)=0 \}$ is a set of roots of the generalized Laguerre polynomial  $L_K^{(a)}(x)$. 
The weight function is given by
\begin{align}
  w_K(a,x) = \frac{(K+a)! \,x}{K! \, (K+1)^2 \, \left(L_{K+1}^{(a)}(x) \right)^2}\ .
  \label{label:weight}
\end{align}
The original Gauss-Laguerre quadrature can be recovered when $a=0$.

We then approximate an integral of a general function $h(x)$ as
\[
\int^{\infty}_{0} dx\ h(x) 
\approx
\sum_{x \in S_K} g_K (a, x) h(x)\ , 
\label{eq:GGLQ}
\]
with $g_K (a, x) = w_K (a, x)\ e^x x^{-a}$.  
The path integral is discretized by employing quadrature rules of different orders for the time-like directions ($x,y$)
and the space-like direction ($z$) as
\[
\int^{\infty}_{0} dx_n dy_n dz_n \ \ \
\Rightarrow \ \ \
\sum_{x_n,y_n \in S_{K_t}}  \sum_{z_n \in S_{K_s}}  g_{K_t}(a,x_n) g_{K_t}(a,y_n) g_{K_s}(a,z_n)   \ . 
\]
Redefining the tensor as
\[
T_{x_i x_j x_k} =
\sqrt{g_{K_t}(a,x_i) g_{K_t}(a,x_j) g_{K_s}(a,x_k)}\ S_{x_i x_j x_k} \ , 
\label{eq:T}
\] 
we can construct a finite-dimensional tensor that approximately reproduces the partition function.
Note that $T$ is a $K_t\times K_t \times K_s$ tensor and symmetric under the interchange of the first two time-like indices.

It would be more convenient to perform the summation over the space-like indices first, which yields the following rank-$4$ tensor:  
\[
T_{ x_i x_j x_k x_l } \vcentcolon =  
\sum_{x_m \in S_{K_s}}
T_{x_i x_k x_m} 
T_{x_j  x_l  x_m} \ ,
\label{eq:T4}
\]
where the indices $x_i$, $x_j$, $x_k$, $x_l$ are all time-like and run from $1$ to $K_t$. 
Accordingly, the partition function can be approximately described by a tensor network consisting of the rank-$4$ tensors: 
\[
Z' 
\approx 
 {\rm Tr} \prod T_{  x_i x_j x_k x_l }\ , 
\label{eq:GLQ4}
\] 
where ${\rm Tr}$ stands for the whole contraction with respect to the indices, 
and $\prod$ the product of all the rank-$4$ tensors in the tensor network. 
\begin{figure}[h]
\centering
\includegraphics[width=6.0in]{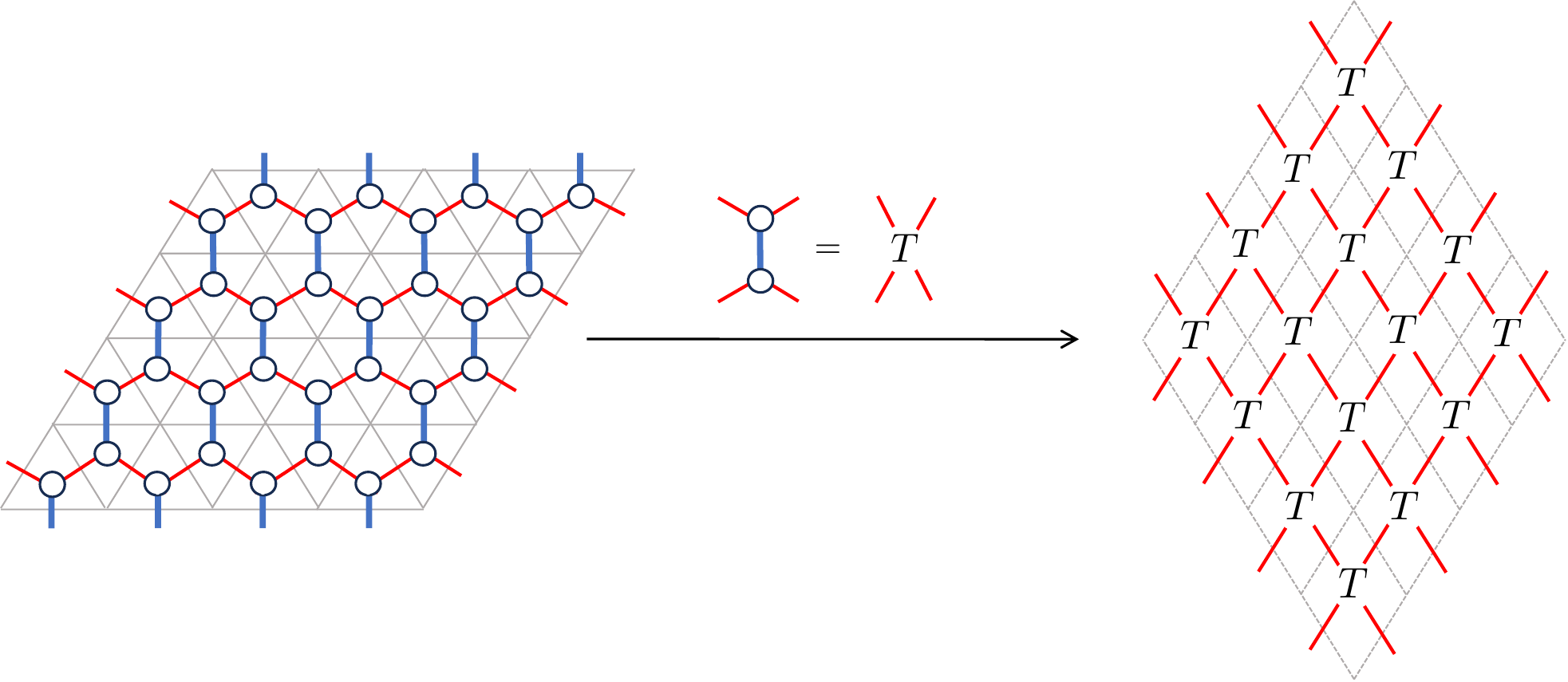}
\caption{The honeycomb tensor network mapped to the square-lattice tensor network.}
\label{fig:squarelattice}
\end{figure} 

Fig.~\ref{fig:squarelattice} illustrates the transformation in Eq.~(\ref{eq:T4}) and the resulting network in Eq. (\ref{eq:GLQ4}) (the right panel). 
The edges of each triangle are discretized through the quadrature rule, 
and in Fig.~\ref{fig:squarelattice} all of them are represented as contracted tensor indices. 
The transformation from the left to the right panel of Fig.~\ref{fig:squarelattice} corresponds to summing over all internal indices associated with the space-like edges (blue links), resulting in a tensor network of rank-$4$ tensors. 
This operation can be carried out without any further truncation.
Note that, while in Fig.~\ref{fig:network} the integrals were associated with the shared links, after Fig.~\ref{fig:squarelattice} these integrals are replaced by finite-dimensional summations.  
As will be seen later in Fig.~\ref{fig:4-8_dual}, the same manipulation is applied there as well.
 
Due to this treatment, the value of the original network is, in principle, obtained by evaluating the network shown in Fig.~\ref{fig:squarelattice} and then taking the limits $K \rightarrow \infty$ and $D \rightarrow \infty$. 
In the actual calculations presented later, we use sufficiently large values of $K_s, K_t$ and $D$ such that the results are converged.

The expectation value of a local observable can also be evaluated using the tensor-network method.
Let us consider four time-like edges (four red links) attached to a single space-like edge (blue link), and denote the corresponding variables by $x_i,x_j,x_k,x_l$.
We then consider an operator ${\cal O} (x_i,x_j,x_k,x_l)$ that depends only on these four links. 
The area $A$ and edge lengths $\ell_e$ are examples of operators belonging to this class.
For operators of this type, the expectation value can be represented by a tensor network that contains a single impurity tensor.
For the discretized network, such a tensor network is given as follows: 
\[
I_{{\cal O}} =
 {\rm Tr}\ \widetilde{T}_{x_i x_j x_k x_l} \prod T_{  x_m x_n x_o x_p }\ , \ \ \ \text{with}\ \ \ 
\widetilde{T}_{x_i x_j x_k k_l} = {\cal O} (x_i,x_j,x_k,x_l)\ T_{x_i x_j x_k x_l}\ ,
 \label{eq:impurity}
\]
where $\widetilde{T}$ is the impurity tensor.
The tensor networks $I_{{\cal O}}$ and $Z^\prime$ are independently evaluated using the tensor renormalization group.
To obtain the expectation value $\langle {{\cal O}} \rangle$, 
one has to evaluate the ratio $I_{{\cal O}}/Z^\prime$ and then take the limits where $K_s$ and $K_t$ are sent to infinity.

We compare the effects of different choices of the generalized quadrature on the computational precision.
In particular, we examine the cases $a=\alpha$ and $a=2\beta/3$, together with the unimproved case, i.e., $a=0$.
The precision is evaluated by the relative error of the expectation value of the area,
\begin{align}
    \Delta A \vcentcolon= \left| 1 - \frac{\langle A_s \rangle}{\frac{3}{2}\alpha +\beta +\frac{3}{2}} \right|\ 
    .
\end{align}

\begin{figure}[H]
    \centering
    \includegraphics[width=7cm]{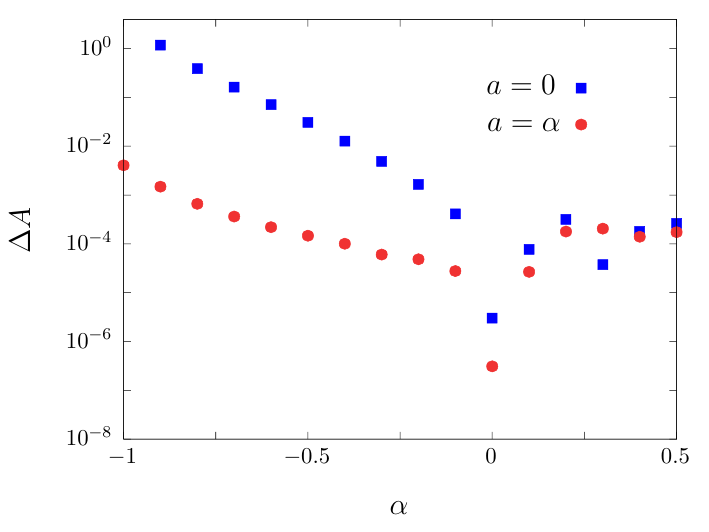}
    \includegraphics[width=7cm]{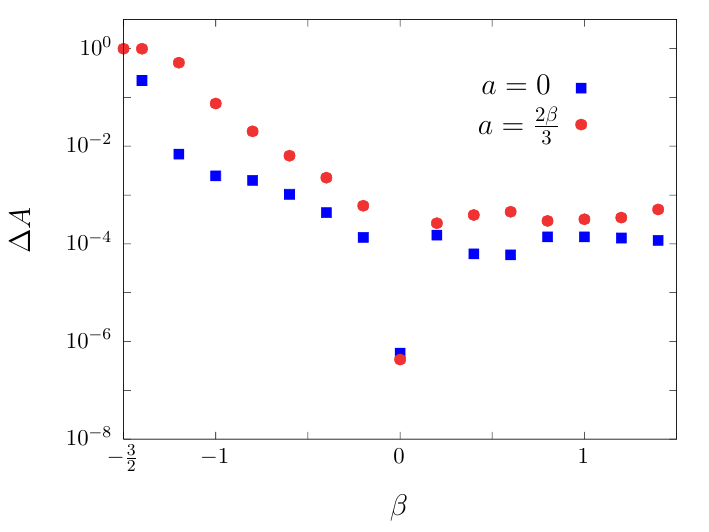}
    \caption{
The $\alpha$-dependence of $\Delta A$ for the $\alpha$-measure (with $\beta=0$) shows that the generalized quadrature significantly improves the precision in the region  $-1<\alpha<0$, whereas the $\beta$-dependence for the $\beta$-measure (with $\alpha=0$) indicates no improvement.
The results are obtained at $N=2\times2^{10}\times2^{10}$, $\beta=0$, $(K_s,K_t)=(200,50)$ and $D=30$.
        } 
    \label{fig:accuracy_A}
\end{figure}

The results are shown in Fig.~\ref{fig:accuracy_A}, 
which compares the values of $\Delta A$ obtained using two different quadrature schemes for the 
$\alpha$-measure ($\alpha\neq 0, \beta=0$) and $\beta$-measure ($\alpha=0, \beta \neq 0$) , respectively. 
The left figure of Fig.~\ref{fig:accuracy_A} indicates that the generalized quadrature with $a=\alpha$ achieves high precision in almost all regions considered.
In contrast, the right figure shows that the generalized quadrature with $a=2\beta/3$ results in poor precision for the $\beta$-measure.
Therefore, in the following analysis, we adopt $a=\alpha$ for the $\alpha$-measure and $a=0$ for the $\beta$-measure.

\subsection{Results for the regular triangulation}

\begin{figure}[H]
    \centering
    \begin{minipage}{0.33\hsize}
      \begin{center}
        \includegraphics[width=5cm]{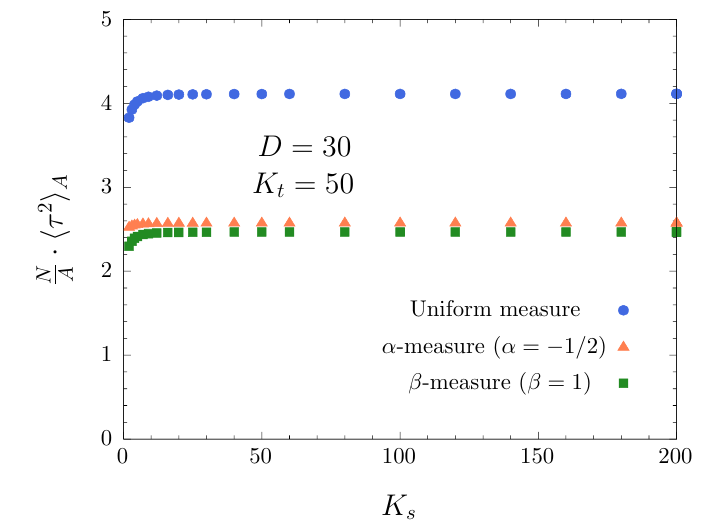}
      \end{center}
    \end{minipage}%
    \begin{minipage}{0.33\hsize}
      \begin{center}
        \includegraphics[width=5cm]{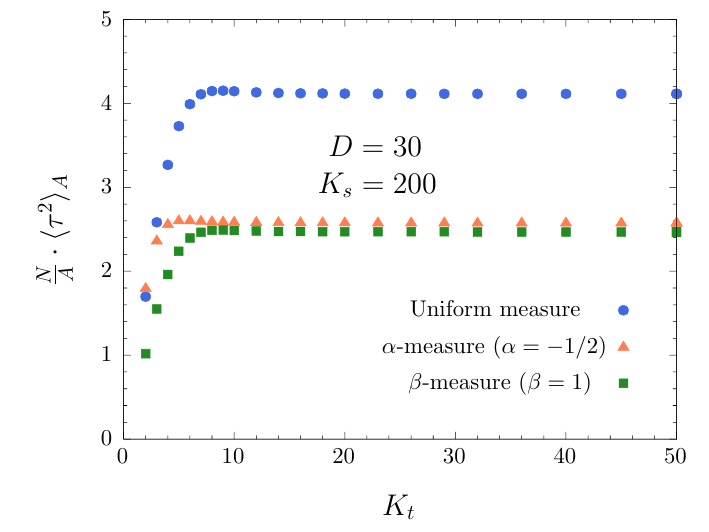}
      \end{center}
    \end{minipage}%
    \begin{minipage}{0.33\hsize}
      \begin{center}
        \includegraphics[width=5cm]{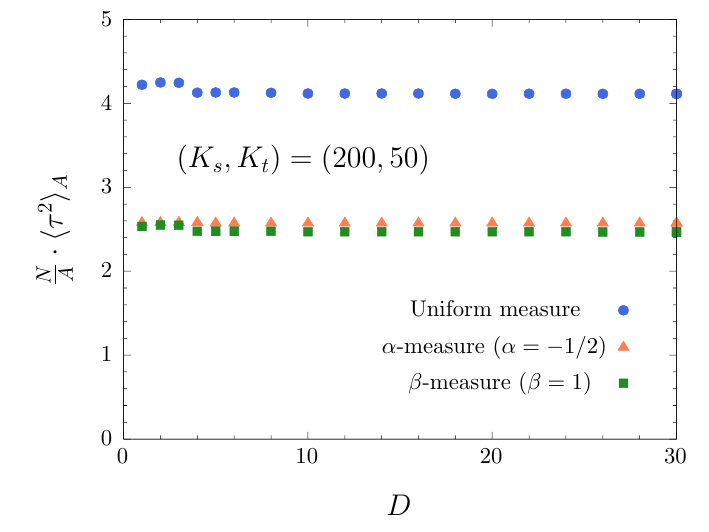}
      \end{center}
    \end{minipage}
    \caption{
The dependence of $\left< \tau^2 \right>_A$ on $K_s$, $K_t$, and $D$
is shown for the uniform measure ($\alpha=\beta=0$), the $\alpha$-measure ($\alpha=-1/2$, $\beta=0$) and the $\beta$-measure ($\alpha=0$, $\beta=1$)
        at $N=2\times2^{10}\times2^{10}$.
        These show the convergence of $\left<\tau^2\right>_A$ for each measure.}
    \label{fig:tau-2_accuracy_6}
\end{figure}

\begin{figure}[H]
    \centering
    \begin{minipage}{0.33\hsize}
      \begin{center}
        \includegraphics[width=5cm]{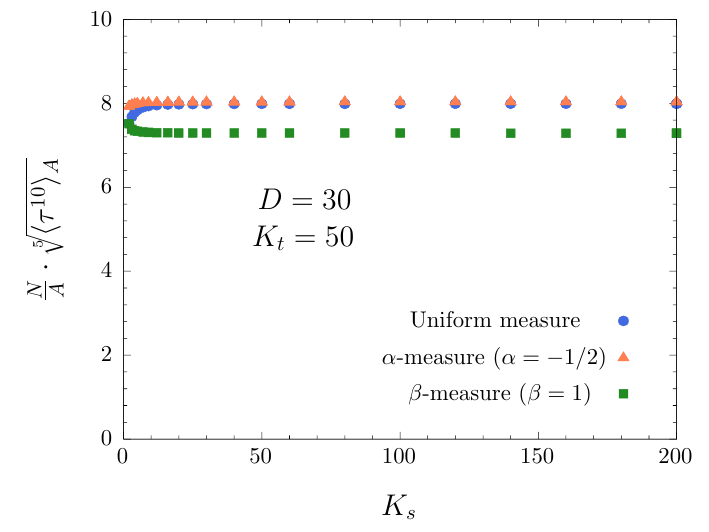}
      \end{center}
    \end{minipage}%
    \begin{minipage}{0.33\hsize}
      \begin{center}
        \includegraphics[width=5cm]{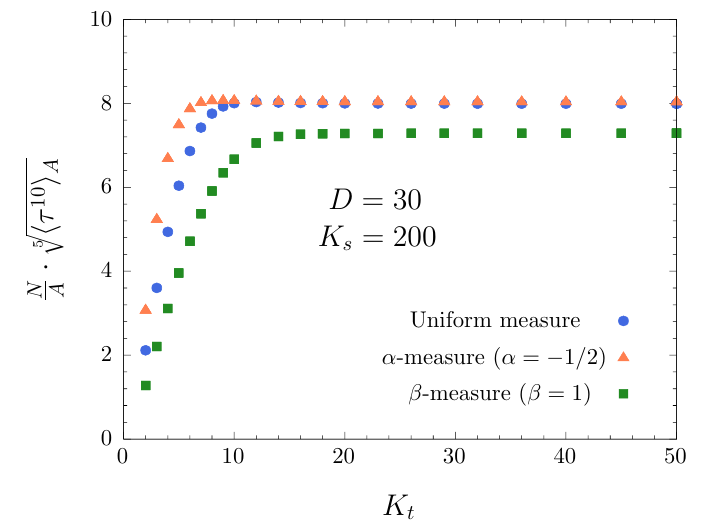}
      \end{center}
    \end{minipage}%
    \begin{minipage}{0.33\hsize}
      \begin{center}
        \includegraphics[width=5cm]{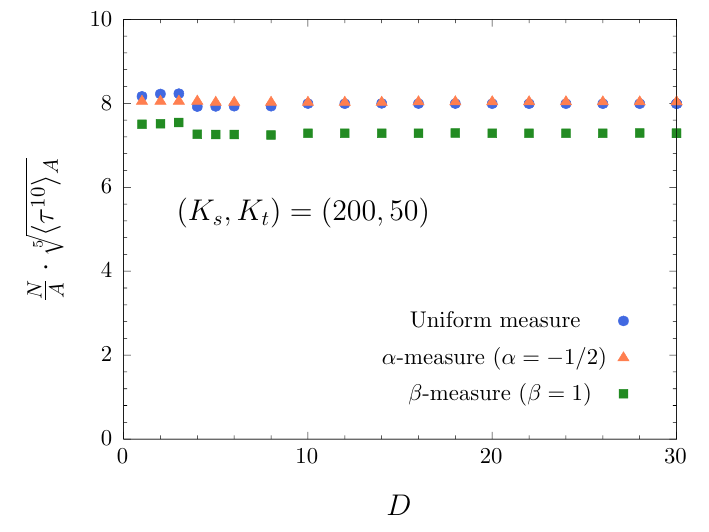}
      \end{center}
    \end{minipage}
    \caption{
The dependence of $\sqrt[5]{\langle \tau^{10} \rangle_A}$ on $K_s$, $K_t$, and $D$
is shown for the uniform measure ($\alpha = \beta = 0$), the $\alpha$-measure ($\alpha = -1/2$, $\beta = 0$), and the $\beta$-measure ($\alpha = 0$, $\beta = 1$)
at $N = 2 \times 2^{10} \times 2^{10}$ and $\mu = 0$.
The figure indicates that $\langle \tau^{10} \rangle_A$ approaches finite values for all the measures.}
    \label{fig:tau-10_accuracy_6}
\end{figure}

We present the results of our investigation into the finiteness of $\left<\tau^n\right>_A$ at large $N$ 
for the regular triangulation (Fig.~\ref{fig:triangulation}).
We first focus on specific values of $\alpha$ and $\beta$, and then examine the dependence on these parameters.
The measure parameters are set to $\alpha = -1/2$ for the $\alpha$-measure ($\beta=0$), and $\beta = 1$ for the $\beta$-measure ($\alpha=0$). 
As confirmed in the previous section, these values provide particularly accurate results for the expectation value of area.  
The convergence of $\left< \tau^2 \right>_A$ with respect to $K_s$, $K_t$ and $D$ is shown in Fig.~\ref{fig:tau-2_accuracy_6}, 
demonstrating the finiteness of $\left< \tau^2 \right>_A$. 
Similarly, the finiteness of $\left< \tau^{10} \right>_A$ is confirmed by the results shown in Fig.~\ref{fig:tau-10_accuracy_6}.

\begin{figure}[H]
    \centering
    \includegraphics[width=8cm]{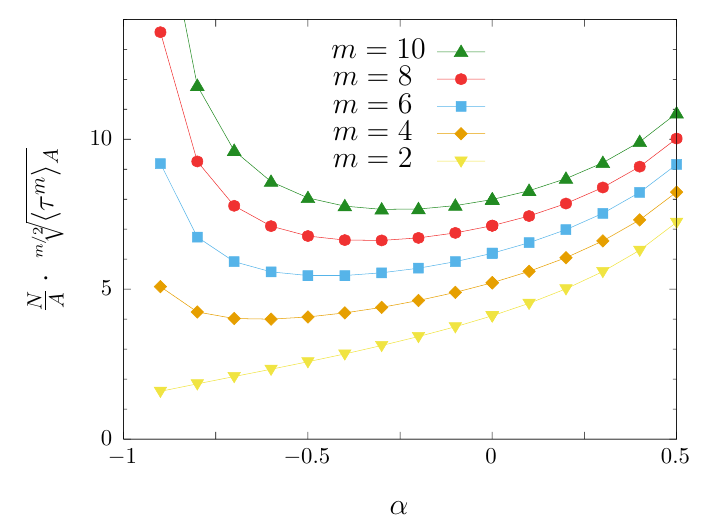}
    \includegraphics[width=8cm]{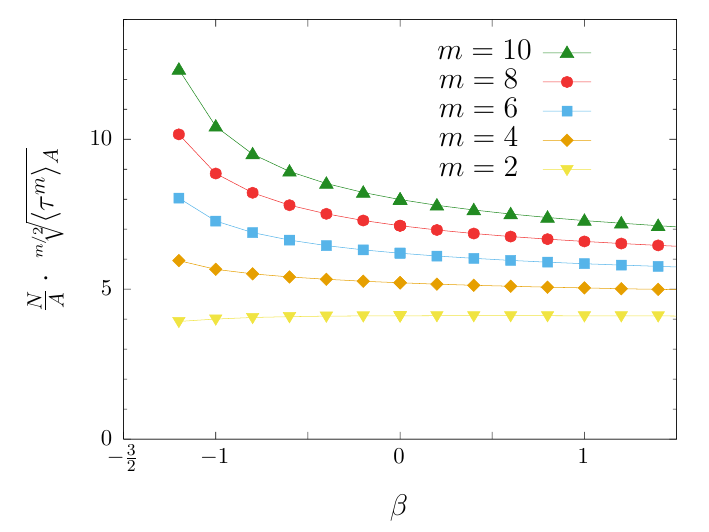}
    \caption{
The $\alpha$ dependence (Left: $\beta = 0$) and $\beta$ dependence (Right: $\alpha = 0$) 
of $\sqrt[m/2]{\langle \tau^{m} \rangle_A}$ for $m = 2, 4, 6, 8, 10$ are plotted 
at $N = 2 \times 2^{10} \times 2^{10}$, $(K_s, K_t) = (200, 50)$, and $D = 30$.
$\langle \tau^{2m} \rangle_A$ behaves as a smooth function of $\alpha$ and $\beta$.}
    \label{fig:tau-alpha-beta_6}
\end{figure}

We then examine the finiteness of $\left< \tau^{n} \right>_A$ for $n=2,4,6,8,10$ while varying the parameters, $\alpha$ and $\beta$. 
The numerical results are shown in Fig.~\ref{fig:tau-alpha-beta_6}, 
which indicate that no peculiar behavior is found as the parameters are varied
and that $\left< \tau^{n} \right>_A$ remains finite in the whole parameter we computed.
As in our previous work, for the case of the regular triangulation, we confirm that pinched geometries are suppressed over all the parameter regions investigated, regardless of the choice of measure. 
Note that approaching the scale-invariant points, $\alpha \to -1$ for the $\alpha$-measure and $\beta \to - \frac{3}{2}$ for the $\beta$-measure, the higher moments with $m=6,8,10$ tend to increase rapidly.


\subsection{Results for the $\mathbf{8}$-$\mathbf{4}$ triangulation}

So far, we have considered a regular triangulation that has a time foliation and a single light cone at each vertex of degree~6 (see Fig.~\ref{fig:triangulation}). 
By numerically checking the finiteness of the higher moments $\langle \tau^n \rangle_A$, we have shown the absence of pinched geometries for that particular triangulation.

\begin{figure}[H]
    \centering
    \includegraphics[width=15.0cm]{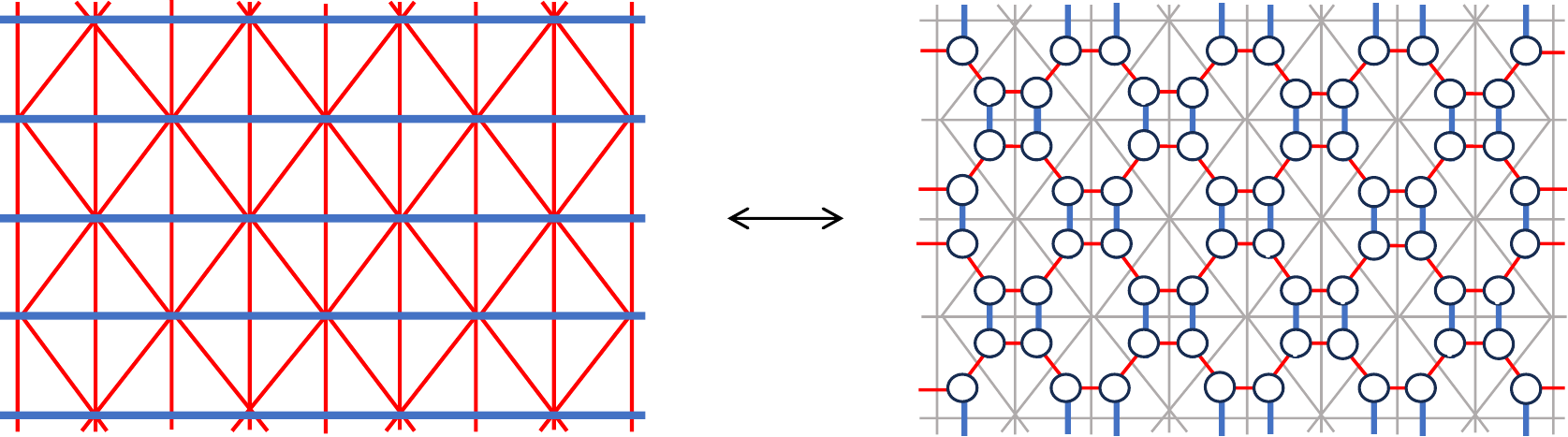}
    \caption{The $8$-$4$ triangulation and the corresponding tensor network.}
    \label{fig:4-8_triangulation}
\end{figure}

In this section, in order to check whether the result above is sensitive to the choice of triangulation, 
we investigate a different type of triangulation, namely, an irregular one depicted in Fig.~\ref{fig:4-8_triangulation}, 
which consists of vertices of degree~8 and~4, and has a time foliation with a single light cone at each vertex. 
We refer to this triangulation as the $8$-$4$ triangulation. 

\begin{figure}[H]
    \centering
    \includegraphics[width=8cm]{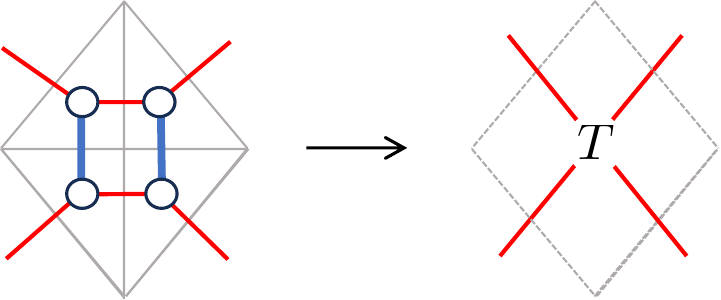}
    \caption{A construction of the rank-$4$ tensor for the $8$-$4$ triangulation.}
    \label{fig:4-8_dual}
\end{figure}

Regarding the $8$-$4$ triangulation, we can construct a rank-4 tensor analogous to Eq.~(\ref{eq:T4}) 
by taking a particular portion of the $8$-$4$ triangulation and partially summing over the indices, 
as illustrated in Fig.~\ref{fig:4-8_dual}. 
Through this procedure, one can again map the tensor network consisting of the rank-$3$ tensors to the square-lattice tensor network. 

\begin{figure}[H]
    \centering
    \begin{minipage}{0.33\hsize}
      \begin{center}
        \includegraphics[width=5cm]{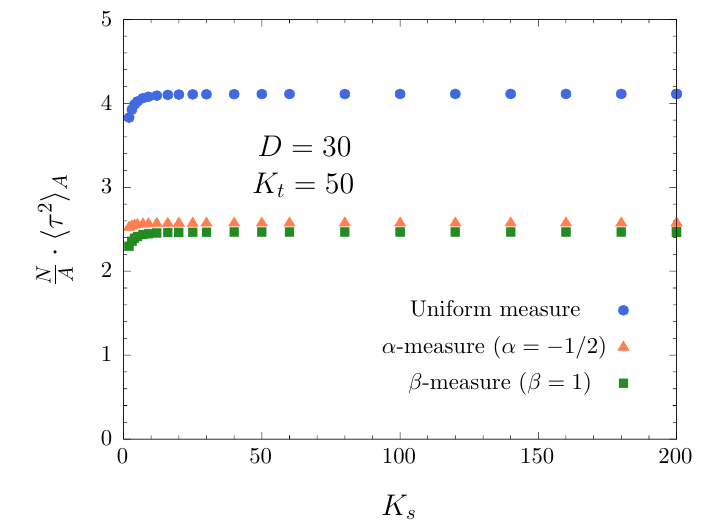}
      \end{center}
    \end{minipage}%
    \begin{minipage}{0.33\hsize}
      \begin{center}
        \includegraphics[width=5cm]{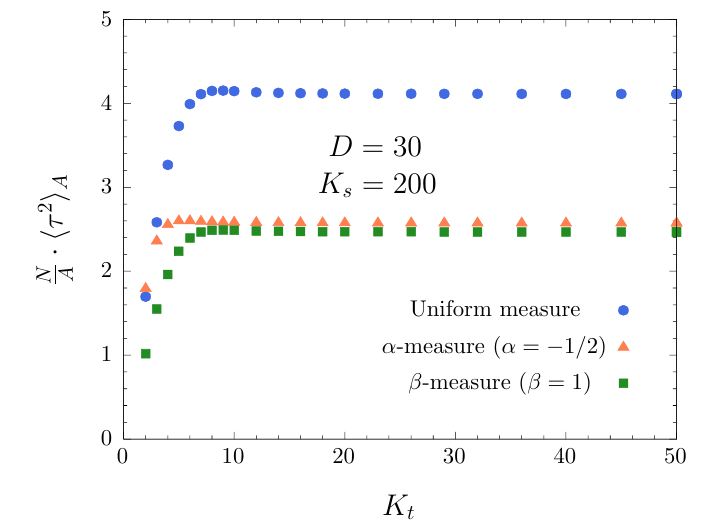}
      \end{center}
    \end{minipage}%
    \begin{minipage}{0.33\hsize}
      \begin{center}
        \includegraphics[width=5cm]{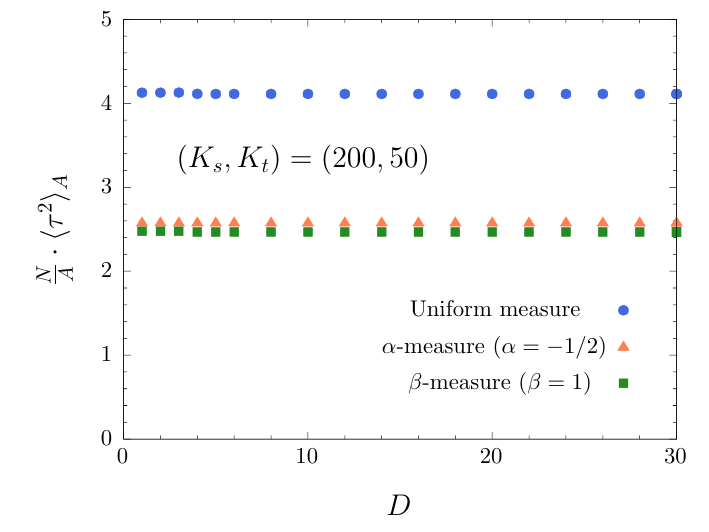}
      \end{center}
    \end{minipage}
    \caption{
The $K_s$, $K_t$, and $D$ dependences of $\langle \tau^2 \rangle_A$
for the $8$-$4$ triangulation are shown
for the uniform measure ($\alpha = \beta = 0$), the $\alpha$-measure ($\alpha = -1/2$, $\beta = 0$), and the $\beta$-measure ($\alpha = 0$, $\beta = 1$),
at $N = 4 \times 2^{10} \times 2^{10}$.}
    \label{fig:tau-2_accuracy_84}
\end{figure}

\begin{figure}[H]
    \centering
    \begin{minipage}{0.33\hsize}
      \begin{center}
        \includegraphics[width=5cm]{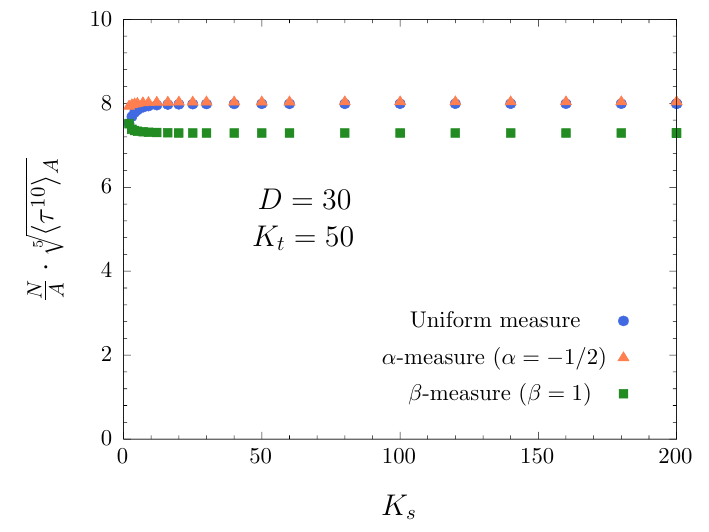}
      \end{center}
    \end{minipage}%
    \begin{minipage}{0.33\hsize}
      \begin{center}
        \includegraphics[width=5cm]{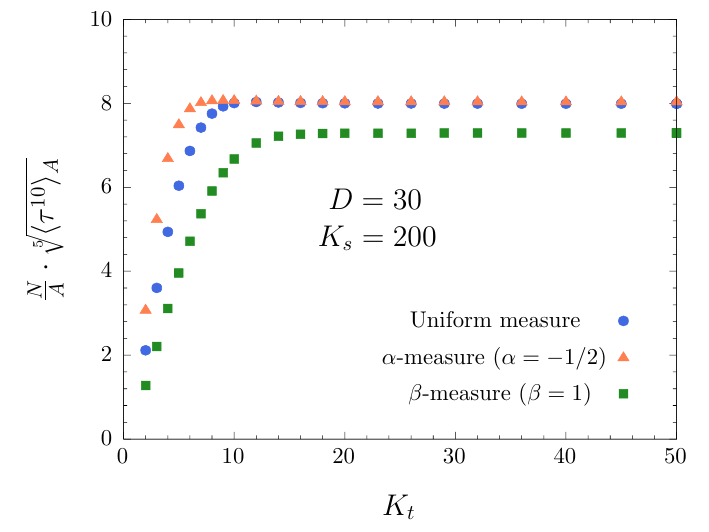}
      \end{center}
    \end{minipage}%
    \begin{minipage}{0.33\hsize}
      \begin{center}
        \includegraphics[width=5cm]{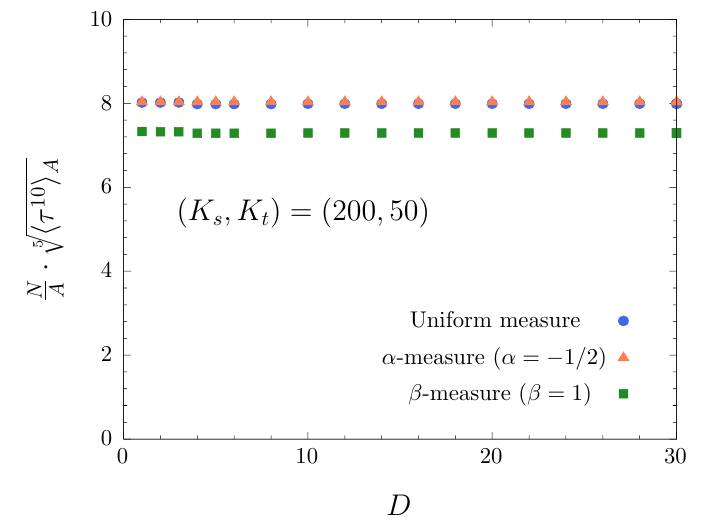}
      \end{center}
    \end{minipage}
    \caption{
        The $K_s$, $K_t$ and $D$ dependences of $\sqrt[5]{ \left< \tau^{10} \right>_A }$
        of the $8$-$4$ triangulation are shown
        for the uniform measure ($\alpha=\beta=0$), the $\alpha$-measure ($\alpha=-1/2$, $\beta=0$) and the $\beta$-measure ($\alpha=0$, $\beta=1$)
        at $N=4\times2^{10}\times2^{10}$. }
    \label{fig:tau-10_accuracy_84}
\end{figure}

Concerning the $\alpha$-measure with $\alpha = -1/2$, the $\beta$-measure with $\beta = 1$, and the uniform measure characterized by $\alpha = \beta = 0$,  
we investigate the $K_s$, $K_t$ and $D$ dependence of $\left< \tau^2 \right>_A$ and $\left< \tau^{10} \right>_A$, 
as shown in Fig.~\ref{fig:tau-2_accuracy_84} and Fig.~\ref{fig:tau-10_accuracy_84}, respectively. 
These figures show that both $\left< \tau^2 \right>_A$ and $\left< \tau^{10} \right>_A$ converge to finite values.

\begin{figure}[H]
    \centering
    \includegraphics[width=8cm]{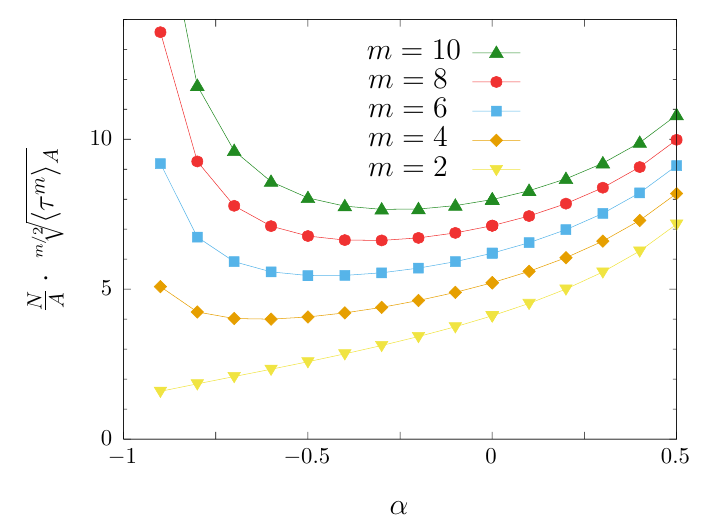}
    \includegraphics[width=8cm]{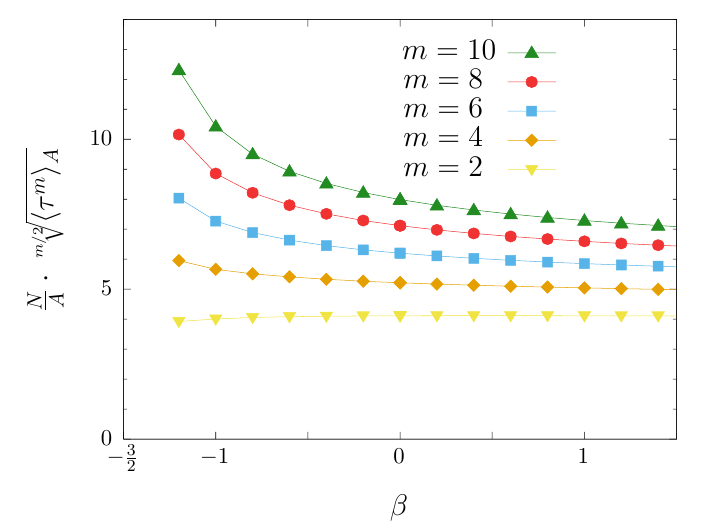}
    \caption{
        The $\alpha$ dependence (Left: $\beta=0$) and $\beta$ dependence (Right: $\alpha=0$) of $\sqrt[m]{\left<\tau^{2m}\right>_A}$ for $m=2,4,6,8,10$ at $N=4\times2^{10}\times2^{10}$, $(K_s,K_t)=(200,50)$ and $D=30$.
As in the case of the regular triangulation, in the $8$-$4$ triangulation the expectation values  $\left<\tau^{2m}\right>_A$ are obtained as smooth and finite functions of $\alpha$ and $\beta$.}
    \label{fig:tau-alpha-beta_84}
\end{figure}

We then explore the $\alpha$ and $\beta$ dependence of the moments 
$\sqrt[m/2]{\left<\tau^{m}\right>_A}$ for $m=2,4,6,8,10$. 
As seen in Fig.~\ref{fig:tau-alpha-beta_84}, no drastic behavior is observed with respect to the variation of the parameters, 
which is similar to the case of the regular triangulation. 
As in the case of regular triangulation, the higher moments with $m=6,8,10$ seem to increase rapidly when approaching the scale-invariant points.

As a result, the suppression of pinched geometries can also be confirmed in the case of the $8$-$4$ triangulation.
This indicates that, within the range investigated, the expectation values of edge lengths remain finite even when the measure and the way of triangulation are varied, suggesting that a smooth geometry can emerge from the Lorentzian model.


\section{Discussions}
\label{sec:Discussions}

We have carefully examined how pinched geometries affect the 2D Lorentzian QRC models when the number of triangles becomes sufficiently large. 
The influence of such geometries can be evaluated through the effective time-like edge length for a fixed total area $A$. 
To see whether this quantity remains finite,
we have numerically analyzed the higher moments of the time-like edge length $\tau$, 
for two types of measures, $l^{2\alpha} dl^2$ and $A^{\beta} dl^2$, that satisfy the inequality $\frac{3}{2}(\alpha + 1) + \beta > 0$, and for two different triangulations.

Across a wide and well-defined range of parameters for the integral measures and different triangulation schemes, we have found that the moments exhibit qualitatively similar behavior, aside from the scale-invariant points. 
All the data consistently show that the effective time-like edge length remains finite with reasonably good accuracy. 
These results strongly suggest that pinched geometries are suppressed and may point toward a certain universality in both the integral measure and the triangulation. 
Since there is currently no straightforward method to construct an integral measure that does not depend on triangulation in QRC, the presence of such universal behavior is a highly desirable property. 
Interestingly, $3$D Lorentzian QRC is an exception, as a discretization-invariant local integral measure has already been constructed, as noted in Ref.~\cite{Borissova:2023izx}\footnote{
In $3$D Lorentzian QRC, the causal structure at vertices was also extensively studied in Ref.~\cite{Asante:2025qbr}.
}. 

The above discussion has been limited to the case of pure gravity. Coupling to matter fields is expected to be quite intriguing, as it could potentially reveal non-trivial phase transitions. 
When matter is coupled, one can fully exploit the advantages of the tensor renormalization group (TRG), as the resulting tensors may remain complex even after a suitable analytic continuation.

In our study, the fundamental building block is a triangle with two time-like edges and one space-like edge. However, alternative triangulation methods can be used, such as those that place a single light-cone at each vertex \cite{Jia:2021xeh}. 
For investigating such models, TRG would be particularly useful.

%%%%%%%%%%%%%%%%%%%%%%%%%%%%%%%%%%%%%%%%%%%%%%%
\section*{Acknowledgement}
We would like to thank 
Jan Ambj\o{}rn, 
Seth Asante, 
Georg Bergner, 
Bianca Dittrich, 
Renata Ferrero, 
Jun Nishimura, 
Kazumasa Okabayashi, 
Sebastian Steinhaus 
and 
Sumati Surya 
for fruitful discussions and encouragements. 
YS is grateful to all the organizers of the workshop ``Quantum Gravity on the Computer 2.0'' at TPI, FSU Jena, for the kind invitation and hospitality.     
YS gratefully acknowledges the kind hospitality of Radboud University, Nijmegen, where part of this work was carried out. 
This work was partially supported by JSPS KAKENHI Grant Number
19K14705, 21K03537, 22H01222, 23H00112, 23K22493, 	25K07318 and JST SPRING Grant Number JPMJSP2125. 
The computation was carried out using the supercomputer ``Flow'' at Information Technology Center, Nagoya University.
% YS's kakenhi: 19K14705, 25K07318
% DK's kakenhi: Fukuma's 23H00112, So's 21K03537, DK's 22H01222 and 23K22493. 
%%%%%%%%%%%%%%%%%%%%%%%%%%%%%%%%%%%%%%%%%%%%%%%%

\section*{Data availability}
The data that support the findings of this article are openly available \cite{Ito:2025}.

\bibliographystyle{ieeetr}

\bibliography{refs}

\end{document}